\documentclass[aps,prl,reprint,showpacs]{revtex4-1}

\usepackage{graphicx}
\usepackage{amsmath, amsthm, amssymb}
\usepackage{textcomp}

\begin{document}

\title{Quantum resonance, Anderson localization and selective manipulations in molecular mixtures by ultrashort laser pulses}

\author{Johannes Flo\ss}
\author{Ilya Sh. Averbukh}
\affiliation{Department of Chemical Physics, The Weizmann Institute of Sciences, Rehovot 76100, ISRAEL}
\date{\today}

\begin{abstract}
We show that the current laser technology used for field-free molecular alignment via impulsive Raman rotational excitation allows for observing long-discussed non-linear quantum phenomena in the dynamics of the periodically kicked rotor.
This includes the scaling of the absorbed energy near the conditions of quantum resonance and Anderson-like localisation in the angular momentum.
Based on this, we show that periodic trains of short laser pulses provide an efficient tool for selective rotational excitation and alignment in a molecular mixture.
We demonstrate the efficiency of this approach by applying it to a mixture of two nitrogen isotopologues ($^{14}$N$_2$ and $^{15}$N$_2$), and show that strong selectivity is possible even at room temperature.
\end{abstract}

\pacs{37.10.Vz, 05.45.Mt, 42.65.Re}


\maketitle


Albeit of its simplicity, the periodically kicked rotor has attracted much attention in the recent decades.
One of the major reasons for the interest in this system is the search for quantum chaos.
In the classical regime, the periodically kicked rotor can exhibit truly chaotic motion.
On the other hand, in the quantum mechanical framework, this chaotic motion is only observed on a short time-scale.
On a longer time-scale, suppression of the chaotic behaviour~\cite{casati79} and Anderson localisation in angular momentum~\cite{fishman82} show up.
Quantum resonance~\cite{casati79,izrailev80} is another distinct feature of the quantum kicked rotor.
If a rotor is kicked at a period that is a rational multiple of the rotational revival time~\cite{robinett04}, its energy grows quadratically with the number of kicks.
Experimentally, the periodically kicked quantum rotor was realised with ultra-cold atoms in periodic potentials~\cite{moore95,raizen99}.
However, the non-discrete character of the atomic momentum hinders the observation of certain phenomena like quantum resonances and chaos assisted tunnelling.
To some degree, this was overcome by using a very narrow initial momentum distribution~\cite{steck01,hensinger01,ryu06}.
Periodically kicked molecules circumvent this problem, since the quantisation of the angular momentum ensures discreteness of the levels.
In the past, it was proposed to use polar molecules interacting with microwave pulses~\cite{bluemel86} for this purpose.
Recently, more proposals~\cite{gong01a,shapiro07} for controlling chaos in polar molecules appeared.
However, due to the complexity of generating the required pulse trains none of these schemes have been realised yet.

In this Letter, we draw attention to the fact that the long-awaited tool~\cite{bluemel86,lee06,spanner04} for exploring the dynamics of the periodically kicked quantum rotor in a molecular system is readily provided by current technology used for laser molecular alignment~\cite{friedrich95,*friedrich95b,stapelfeldt03}.
Here, short laser pulses induce molecular alignment via impulsive Raman-type rotational excitation.
The electric field of the pulse induces anisotropic molecular polarisation, interacts with it, and tends to align the molecular axis along the laser polarisation direction.
An ultra-short laser pulse acts like a kick, and the alignment is observed under field-free conditions after the pulse~\cite{ortigoso99,stapelfeldt03,seideman99a,underwood05}.
Recently, a periodic train of eight pulses was used for inducing enhanced molecular alignment by repeated kicking under the condition of quantum resonance~\cite{cryan09}.
We show that experimental techniques similar to the one used by Cryan~\textit{et al.}~\cite{cryan09} finally allow for observing for the first time effects like scaling of the absorbed energy near the quantum resonance~\cite{wimberger03,*wimberger04} and Anderson localisation~\cite{fishman82} in a real rotational system, a diatomic molecule.
Moreover, we show that these phenomena provide a new toolbox for selective rotational laser manipulations in molecular mixtures~\cite{renard04,fleischer06,*fleischer07}.
To illustrate, we propose and discuss a new scheme for selective alignment of molecular isotopologues, using the aforementioned effects.


We consider interaction of linear molecules with a periodic train of linearly polarised laser pulses.
The pulses are far off-resonant from any molecular transition.
The interaction potential is given as
\begin{equation}
V=-\frac{1}{4}\Delta\alpha \cos^2\theta \sum_{n=0}^{N-1}\mathcal{E}^2(t-n\tau)\,.
\label{potential}
\end{equation}
Here, $\mathcal{E}(t)$ is the temporal envelope of a single laser pulse, $\Delta\alpha=\alpha_{\parallel}-\alpha_{\perp}$, where $\alpha_{\parallel}$ is the polarisability along the molecular axis and $\alpha_{\perp}$ is the polarisability perpendicular to it, $N$ is the number of pulses, $\tau$ is the period of the train, and $\theta$ is the angle between the laser polarisation axis and the molecular axis.
The pulses are assumed to be much shorter than the rotational time-scale of the molecules, so we treat them as $\delta$-kicks.
We introduce the effective interaction strength $P=\Delta\alpha/(4\hbar)\int \mathcal{E}^2(t) dt$.
It reflects the typical change of the molecular angular momentum (in units of $\hbar$) induced by a single laser pulse.
We measure the energy in units of $2B$ and the time in units of $\hbar/(2B)$, where $B=\hbar^2/(2I)$ is the rotational constant and $I$ is the moment of inertia.
The wave function right after interaction with a single pulse is given as
\begin{equation}
|\Psi(t^+)\rangle=e^{iP\cos^2\theta}|\Psi(t^-)\rangle \,,
\end{equation}
where $t^+$ and $t^-$ are the time instants just after the pulse and before it, respectively.
We expand the wave function in the eigenfunctions of the free rotating molecule:
\begin{equation}
|\Psi(t)\rangle=\sum_{J,M} c_{J,M} e^{-iE_Jt} |J,M\rangle\,.
\label{wavepacket}
\end{equation}
We include the centrifugal distortion term in the rotational levels $E_J=J(J+1)/2-DJ^2(J+1)^2/(2B)$, where $D$ is the centrifugal distortion constant.
In the rigid rotor approximation, the frequencies of rotational transitions form an equidistant series with the spacing of $2hB$.
Therefore, the wave packet~\eqref{wavepacket} revives after the revival time $t_{rev}=2\pi$ (in units of $\hbar/(2B)$).
Due to the centrifugal distortion, this revival is not exact.
The transformation of the $c_{J,M}$ coefficients due to a kick is obtained by the numerical procedure described in~\cite{fleischer09}.
Thermal averaging over initial molecular states is done to account for thermal effects.
For symmetric molecules like N$_2$ we take into account the effects of nuclear spin statistics~\cite{herzbergBook}.

We start with considering rotational excitation in the close proximity of the quantum resonance, i.e. $\tau=2\pi+\epsilon$, where $|\epsilon|$ is small.
We adopt here an approach used previously for obtaining a scaling law for the resonances of cold atoms in a pulsed optical standing wave~\cite{wimberger03,*wimberger04}.
For simplicity, we restrict ourselves here to the case of a rigid rotor at zero temperature.
The one-period evolution operator is
\begin{equation}
\hat U = e^{iP\cos^2\theta}e^{-i\epsilon \hat J^2 /2} \,.
\label{floquet}
\end{equation}
Here, $\hat J$ is the angular momentum operator.
We scale the interaction strength as $\tilde P = |\epsilon|P$ and the angular momentum as $\hat I = |\epsilon|\hat J$ and obtain
\begin{equation}
\hat U = e^{i\tilde P\cos^2\theta/|\epsilon|}e^{-i \hat I^2/(\pm2|\epsilon|)} \,,
\label{scaledFloquet}
\end{equation}
where $\pm$ is the sign of $\epsilon$.
In Eq.~(\ref{scaledFloquet}), $|\epsilon|$ can be regarded as an effective Planck's constant~\cite{wimberger03,*wimberger04}.
Therefore, if $|\epsilon|\ll P$, the quantum dynamics of~(\ref{scaledFloquet}) can be described by an effective classical mapping.
It is given as
\begin{align}
\theta_{n+1} =&\, \theta_n \pm I_n &
I_{n+1} =&\, I_n + \tilde P \sin(2\theta_n) \,.
\label{mapping}
\end{align}
Since we assume zero temperature here, no rotations around the $z$-axis are excited.
If the change of the angle $\theta$ with each pulse is much smaller than $2\pi$ (i.e. when the kick strength is sufficiently low), we can make use of the pendulum approximation~\cite{lichtenberg92} to write the classical mapping~(\ref{mapping}) as differential equations:
\begin{align}
d\theta/dn=&\,\pm I &
d I/dn=&\,-\tilde P \sin(2\theta) \,.
\label{eom}
\end{align}
By rescaling the angular momentum to $\tilde I = \sqrt{|\epsilon|P}I$ and the time to $x=\sqrt{|\epsilon|P}n$, Eqs.~(\ref{eom}) become parameterless, apart from the sign of $\epsilon$.
We obtain $\tilde I(x)$ via the Jacobi elliptic functions.
The final rotational energy is given as
\begin{equation}
E(\epsilon,P,N)=\frac{\tilde I_{\pm}^2(x) P}{2|\epsilon|} \,,
\label{eq10}
\end{equation}
where $\pm$ indicates the dependence on the sign of $\epsilon$.
We average over the initial angle $\theta_0$, which has the distribution $f(\theta_0)=0.5\sin(\theta_0)$.
The averaged value of the absorbed energy is normalised by the energy absorbed in the case of resonant kicking, which is
\begin{equation}
\frac{1}{2}\langle 0| e^{-iNP\cos^2\theta} \hat J^2 e^{iNP\cos^2\theta} | 0\rangle = \frac{4}{15} N^2P^2 \,.
\end{equation}
The scaled average energy is therefore given by
\begin{equation}
R(\epsilon,P,N)=\frac{\langle \tilde I_{\pm}^2(x) \rangle P}{2|\epsilon|}\frac{15}{4 N^2P^2} = \frac{15}{8}\frac{\langle \tilde I_{\pm}^2(x) \rangle}{x^2} \,.
\label{scaledenergy}
\end{equation}
Here, the angular brackets denote averaging over $\theta_0$.
As can be seen, the scaled energy is a function of the combined scaling variable $x=\sqrt{|\epsilon| P}N$ alone, and not of the three parameters $\epsilon$, $P$ and $N$ independently.
A similar result was previously obtained for the kicked cold atoms~\cite{wimberger03,wimberger04} with the same scaling variable but a different scaled energy $R(x)$.
\begin{figure}
\includegraphics{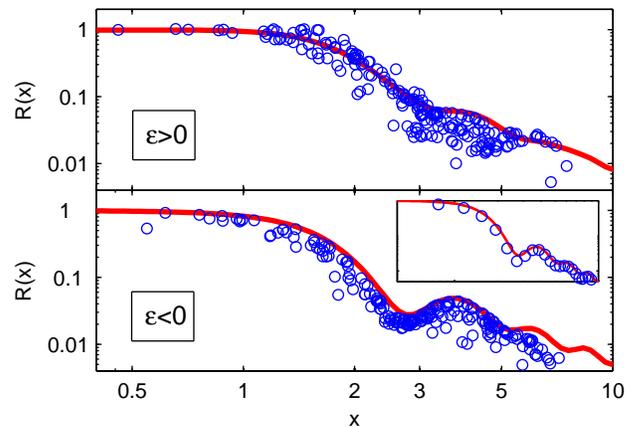}
\caption{\label{scalingcomparison}(Color online) Scaled energy $R(x)$ calculated by Eq.~\eqref{scaledenergy} (solid line) and by direct quantum-mechanical simulations (circles).
The latter are done for $^{15}$N$_2$ at $T=10~\text{K}$, for random values of the interaction parameters $\epsilon$, $P$ and $N$ in the intervals $|\epsilon| \in [0.0001\pi,0.025\pi]$, $N \in [4,40]$, with the total interaction strength $P_{tot}=NP$ limited to $4\le P_{tot}\le24$.
The inset shows the data corresponding to a pulse train with $\epsilon=-0.05$ and $P=3$, for which the absolute energy is shown in Fig.~\ref{energyFig} (diamonds).
}
\end{figure}
In Fig.~\ref{scalingcomparison} the scaled energy obtained by the effective classical mapping is compared to exact quantum mechanical calculations for non-rigid $^{15}$N$_2$ at a temperature of $T=10~\text{K}$.
The qualitative agreement between both is rather good, despite the fact that the effective classical result was derived for a rigid rotor at zero temperature.

We now present the results of simulations for the rotational excitation of the isotopologue $^{15}$N$_2$, interacting with a periodic train of identical laser pulses.
We show two cases.
Firstly, we consider excitation at quantum resonance ($\epsilon=0$) and near it ($|\epsilon| \ll 1$), i.e. in the regime of the scaling law~\eqref{scaledenergy}.
Secondly, we simulate far off-resonant excitation in the regime of Anderson localisation.
For the latter scenario we consider $\epsilon=-0.44$, as it represents the difference in the revival times of $^{14}$N$_2$ and $^{15}$N$_2$.
The kick strength is set to $P=3$, which is comparable to the pulses used in experiment~\cite{cryan09}.
In the simulations, we account for centrifugal distortion and thermal effects, with temperature $T=10~\text{K}$.

\begin{figure}
\includegraphics[width=\linewidth]{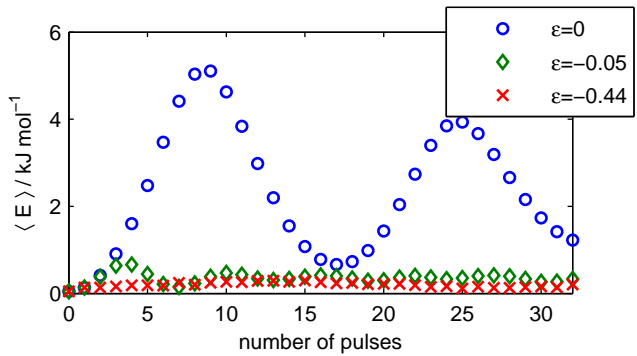}
\caption{\label{energyFig}(Color online) Absorbed rotational energy of $^{15}$N$_2$ at $T=10~\text{K}$ after interacting with a periodic train of ultra-short laser pulses with $P=3$.
The results are shown for different detunings $\epsilon$.}
\end{figure}

In Fig.~\ref{energyFig} the absorbed energy is shown as a function of the number of pulses $N$.
As expected, for kicking on resonance (circles) the absorbed energy grows initially as the square of $N$.
At $N\approx8$, this growth saturates and eventually the energy starts oscillating, which is due to the centrifugal distortion.
For the near-resonant case (diamonds), the absorbed energy follows the predictions by the scaling law.
This is also shown by the inset in Fig.~\ref{scalingcomparison}.
For far off-resonant kicking (crosses), only a very weak absorption of energy is observed.

\begin{figure}
\includegraphics[width=\linewidth]{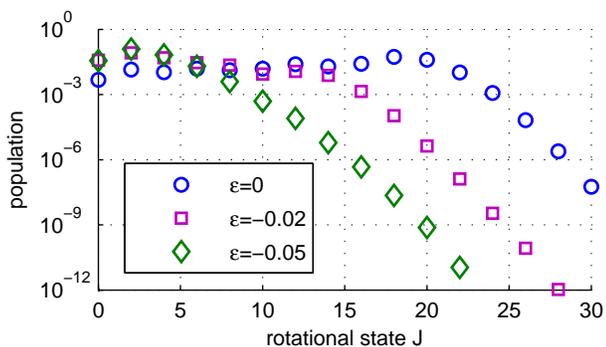}
\caption{\label{resonance}(Color online) Population of the rotational levels of $^{15}$N$_2$ at $T=10~\text{K}$ after interaction with a train of eight pulses with $P=3$, for resonant ($\epsilon=0$) and near-resonant ($|\epsilon|\ll 1$) kicking.}
\end{figure}
In Fig.~\ref{resonance} the population of the rotational levels $J$ after a train of eight pulses, tuned on or close to the quantum resonance, is shown.
Due to the selection rules of the Raman type excitation ($\Delta J=0,\pm2$), the sub-sets of even and odd rotational levels evolve independently, so for clarity we show only the even ones.
The population of the odd levels behaves similarly, with a correction by a constant factor arising from nuclear spin statistics~\cite{herzbergBook}.
Under the condition of exact resonance, the distribution is divided into a flat (in logarithmic scale) plateau region, and a fast decay after some cut-off value of $J$.
The cut-off marks the maximum angular momentum supplied by $N$ pulses to a classical rotor (for a rigid rotor this is $J= N P$), and the fast decay is due to the ``tunnelling'' into the classically forbidden region.
When the detuning is increased, one can see a monotonous deformation of the population curve, whilst the general shape (plateau and fast decay after a cut-off) remains intact.

\begin{figure}
\includegraphics[width=\linewidth]{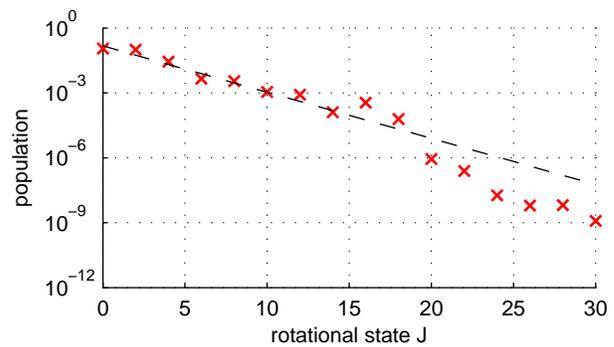}
\caption{\label{Anderson}(Color online) Population of the rotational levels of $^{15}$N$_2$ at $T=10~\text{K}$ after interaction with a train of eight pulses with $P=3$, for an off-resonant pulse train ($\epsilon=-0.44$).
The dashed line indicates the exponential decay caused by Anderson-like localisation.}
\end{figure}
For larger detuning -- shown in Fig.~\ref{Anderson} --, the distribution is completely different  and reflects Anderson localisation in the system.
Instead of the plateau, a clear linear (in logarithmic scale) decay over several orders of magnitude is seen, starting from $J=0$.
We have verified by additional simulations that this part of the distribution is quite insensitive to variations of the detuning, in contrast to the case of near-resonance kicking (Fig.~\ref{resonance}).
This is an intrinsic property of Anderson localisation in the periodically kicked rotor~\cite{fishman82}.
The deviations from linearity for $J\gtrsim20$ are mainly due to the finite number of pulses.

The dependence of the rotational excitation on the detuning $\epsilon$ can be used for isotope-selective rotational excitation, as reflected in transient beats of molecular alignment.
The alignment is quantified by the expectation value of $\cos^2\theta$ (alignment factor).
As an example, we consider the two nitrogen isotopologues $^{14}$N$_2$ and $^{15}$N$_2$, with the rotational revival times $8.35~\text{ps}$ and $8.98~\text{ps}$, respectively.
If the pulse train is resonant to $^{14}$N$_2$, the detuning for $^{15}$N$_2$ is $\epsilon=-0.44$, which corresponds to the far off-resonant case considered above.

\begin{figure}
\includegraphics{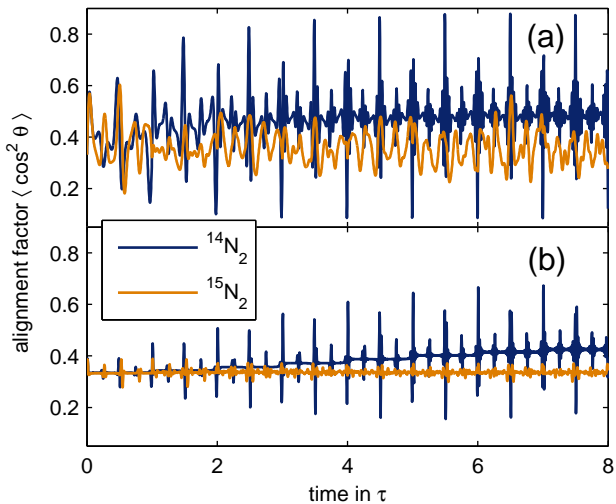}
\caption{\label{alignment}(Color online) Alignment factors of $^{14}$N$_2$ and $^{15}$N$_2$ molecules, interacting with a train of laser pulses.
Panel~(a) shows the results for $T=10~\text{K}$, panel~(b) for $T=298~\text{K}$.
The pulses are applied with a constant time-delay $\tau$, which is equal to the rotational revival time of $^{14}$N$_2$ (8.35~ps for $T=10~\text{K}$, 8.37~ps for $T=298~\text{K}$).}
\end{figure}

In Fig.~\ref{alignment}~(a) we show the alignment signal for the two isotopologues interacting with a train of eight pulses.
For the resonantly kicked $^{14}$N$_2$, strong alignment builds up, in agreement with~\cite{cryan09}.
On the other hand, for $^{15}$N$_2$ the alignment signal is considerably weaker, and therefore selective alignment is achieved.

A remarkable effect can be seen in Fig.~\ref{alignment}~(b), showing the results at $T=298~\text{K}$.
Whilst at $T=10~\text{K}$ the difference between the peak alignment and the isotropic value ($1/3$) is three times larger for $^{14}$N$_2$ than for $^{15}$N$_2$, at $T=298~\text{K}$ it is even six times larger for $^{14}$N$_2$.
Therefore, the selectivity of the alignment process is improved at high temperature.
The reason is that the resonant species constructively accumulates the effect of the kicking pulses even at high temperature under the conditions of quantum resonance~\cite{cryan09}, whereas the coherent excitation of the off-resonant isotopologue is hindered strongly due to the excitation randomisation caused by fast thermal rotation between subsequent kicks.


Summarising, we showed that the current laser technology used for non-adiabatic field-free molecular alignment via impulsive Raman-type rotational excitation is sufficient for observing several fundamental phenomena in the dynamics of the periodically kicked quantum rotor.
We derived a scaling law for the energy absorption for a molecule kicked close to the quantum resonance, similar to the one that was previously discussed in the context of atom optics~\cite{wimberger03,wimberger04}.
Moreover, we showed that Anderson-like localisation can be observed in diatomic molecules, using experimentally feasible pulse trains.

We applied the discussed phenomena to the challenging task of selective rotational excitation in a molecular mixture.
A new scheme for isotopologue selective alignment, which is based on Anderson localisation in periodically kicked systems, is proposed.

The results of this work may stimulate experimental observations of quantum dynamical effects in the periodically kicked quantum rotor, using for the first time a real rotational system, a molecule.
Moreover, our results may pave the way to novel methods for selective manipulation in mixtures of molecular species by means of the above mentioned phenomena.

After submission of this paper, an experimental work by Milner~\textit{et al.}~\cite{zhdanovich12} appeared, in which a train of  about six periodic (although not identical) pulses was used in a scheme along the lines of the present Letter.
By changing the spacing between the pulses, isotopologue selective rotational excitation in a mixture of $^{14}$N$_2$ and $^{15}$N$_2$ was achieved.
Moreover, by scanning the train period around a fractional revival, Milner~\textit{et al.} achieved selective rotational excitation of different nuclear spin isomers of $^{15}$N$_2$.

We thank Shmuel Fishman, Sergey Zhdanovich and Valery Milner for fruitful discussions.
Financial support of this research by the ISF and the DFG is gratefully acknowledged.
The work of JF is supported by the Minerva Foundation.
This research is made possible in part by the historic generosity of the Harold Perlman Family.


%

\end{document}